\documentstyle[preprint,aps,epsfig,epsf,rotate]{revtex}

\begin{document}
\tighten

\preprint{
\vbox{ 
\hbox{ADP-01-07/T442}
\hbox{JLAB-THY-01-09}
\hbox{CTP-3092}
}}

\title{Chiral Extrapolation of Lattice Moments of Proton Quark
	Distributions}

\author{W.~Detmold$^1$, W.~Melnitchouk$^{1,2}$, J.W.~Negele$^3$,
	D.B.~Renner$^3$ and A.W.~Thomas$^1$}
\address{$^1$ Special Research Centre for the Subatomic Structure of
	Matter, and Department of Physics and Mathematical Physics,
	Adelaide University, 5005, Australia}
\address{$^2$ Jefferson Lab, 12000 Jefferson Avenue,
	Newport News, VA 23606}
\address{$^3$ Center for Theoretical Physics, Massachusetts Institute of
	Technology, Cambridge, MA 02139}

\maketitle

\begin{abstract}
We present the resolution of a long-standing discrepancy between the
moments of parton distributions calculated from lattice QCD and their
experimental values.
We propose a simple extrapolation formula for the moments of the
nonsinglet quark distribution $u - d$, as a function of quark mass,
which embodies the general constraints imposed by the chiral symmetry
of QCD.
The inclusion of the leading nonanalytic behavior leads to an excellent
description of both the lattice data and the experimental values of the
moments.
\end{abstract}

\newpage

Although historically, deep-inelastic scattering from the nucleon
provided an important test of perturbative QCD, precision measurements of
parton distribution functions (PDF's) in these experiments now provide
crucial, fundamental information about the nonperturbative structure of
the nucleon.
Recent discoveries which have had a profound impact on our understanding
include the proton spin crisis \cite{EMC}, the Gottfried sum rule
violation \cite{NMC}, and to a certain extent the nuclear EMC effect
\cite{NUCEMC}.
Future experiments aimed at testing whether $\Delta \bar{u}$ and
$\Delta \bar{d}$ are equal or whether $s(x)$ differs from $\bar{s}(x)$
should also serve to deepen our understanding of the nonperturbative
origin of parton distributions.

A decade or more of rigorous, nonperturbative calculations of the moments
of PDFs in the nucleon within lattice QCD has so far led to a major
impasse.
The values of the first three nontrivial moments typically lie some 50\%
above the corresponding experimental data.
Since PDF moments are benchmark calculations of hadron structure in
lattice QCD, an unresolved discrepancy of this order of magnitude in such
fundamental quantities would seriously undermine the credibility of any
{\it ab initio} calculation of hadronic properties, and therefore
represents a crucial challenge in hadronic physics.
In this Letter we explain for the first time the physics required to
resolve this problem.
We show that inclusion of the nonanalytic chiral behavior of the moments
of $u - d$ as a function of quark mass removes the discrepancy.

At first sight, when one thinks of structure functions in terms of
light-cone correlation functions of currents measured at high energy and
momentum transfer, it may appear remarkable that spontaneous chiral
symmetry breaking and the associated pion cloud could play an essential
quantitative role. Indeed, the 50\% effects may seem all the more
puzzling when chiral corrections to lattice hadron mass calculations are
known to be far smaller. However, as discussed below, moments of structure
functions correspond to matrix elements of local operators in the hadron
ground state. Furthermore, by the familiar variational principle in
quantum mechanics, a first order error in the wave function yields a
first-order error in matrix elements of general operators while producing
only a second order error in the energy, so much larger errors in
operators than masses should be expected.

Lattice calculations of parton distributions in Euclidean space-time are
based on the operator product expansion --- one calculates the matrix
elements of certain local operators.
The results are directly related to the moments of the measured PDFs:
\begin{eqnarray}
\langle x^n \rangle_q
&=& \int_0^1 dx\ x^n\
    \left( q(x,Q^2) + (-1)^{n+1} \bar q(x,Q^2) \right)\ ,
\label{eq:mom}
\end{eqnarray}
where the distribution $q(x,Q^2)$ is a function of the Bjorken variable
$x$ and the momentum scale $Q^2$.
The operator product expansion relates the moments $\langle x^n \rangle_q$
to forward nucleon matrix elements of local twist-2 operators, which for
nonsinglet distributions are given by\ 
${\cal O}_{\{ \mu_1 \ldots \mu_{n+1} \}}
= \overline \psi\ \gamma_{\{ \mu_1}
  \stackrel{\leftrightarrow}{D}_{\mu_2} \ldots 
  \stackrel{\leftrightarrow}{D}_{\mu_{n+1} \}} \psi$,
where $\psi$ is the quark field, $D_\mu$ the covariant derivative and
$\{\ldots\}$ represents symmetrization of the Lorentz indices.
As a result of operator mixing on the lattice, all lattice calculations
have so far been restricted to $n \leq 3$.
Nevertheless, many features of the PDFs can be reconstructed from just
the lowest few moments \cite{WEIGL}.

Early calculations of structure functions within lattice QCD were
performed by Martinelli and Sachrajda \cite{MS}.
The data used in the present analysis, shown in Fig.~1 for the
$n=1$, 2 and 3 moments of the $u - d$ difference in the
$\overline{\rm MS}$ scheme, are taken from the more recent and extensive
calculations by the QCDSF \cite{QCDSF96,QCDSF97,BEST} and MIT \cite{MIT}
groups.
The data include results from quenched simulations at $\beta=6.0$ for
several values of $\kappa$, which for a world average lattice spacing of
$a = 0.1$~fm correspond to quark masses ranging from 30 to 190~MeV.
In addition, we include unquenched data from the MIT group, which has also
performed the first full QCD calculations at $\beta=5.6$ (corresponding to
the same $a$ as for the quenched data with $\beta=6.0$) using the SESAM
configurations \cite{SESAM}.
The unquenched results are consistent with the quenched data, indicating
that internal quark loops do not appear to play an important role at the
quark masses considered.
Rather than show the moments versus the scale and renormalization scheme
dependent quark mass, we plot the data as a function of the pion mass
squared, $m_\pi^2 \propto m_q$.
For the $n=1$ moment we retain only the data corresponding to the
statistically most accurately determined operator
${\cal O}_{44} - 1/3 \sum_{i=1}^{3} {\cal O}_{ii}$
\cite{QCDSF96,QCDSF97,BEST,MIT}.
To avoid finite volume effects \cite{FINITEV}, we exclude points at the
lowest quark masses from the data sets of Refs.~\cite{QCDSF97} and
\cite{MIT}.
The moments correspond to a momentum scale of
$Q^2 = 1/a^2 \approx 4$~GeV$^2$.

Note that matrix elements of the operators
${\cal O}_{\{ \mu_1 \ldots \mu_{n+1} \}}$ include both connected and
disconnected diagrams, corresponding to operator insertions in quark
lines which are connected or disconnected (except through gluon lines)
with the nucleon source.
Since the evaluation of disconnected diagrams is considerably more
difficult numerically, only exploratory studies of these have been
completed \cite{DISCON} and the present work will treat only connected
diagrams.
However, because the disconnected contributions are flavor independent
(for equal $u$ and $d$ quark masses), they cancel exactly in the
{\em difference} of $u$ and $d$ moments.
Therefore it is appropriate to compare connected contributions to lattice
$u-d$ moments with moments of phenomenological PDFs \cite{PARAMS}.

To compare the lattice results with the experimentally measured moments,
one must extrapolate the data from the lowest quark mass used
($\sim 50$~MeV) to the physical value ($\sim$ 5--6~MeV).
Naively one extrapolates to the physical quark mass assuming that the
moments depend linearly on the quark mass.
However, as shown in Fig.~1 (long dashed lines), a linear extrapolation
of the world lattice data for the $u-d$ moments overestimates the
experimental values by some 50\% in all cases.
This suggests that important physics is still being omitted from the
lattice calculations and their extrapolations.
It is crucial, if one is to have confidence in lattice calculations of
hadronic observables, that the origin of this discrepancy is identified.

Indeed, one knows on very general grounds that a linear extrapolation
in $m_q \sim m_\pi^2$ must fail because it omits the crucial
nonanalytic structure associated with chiral symmetry breaking.
Even at the lowest quark mass accessed on the lattice, the pion mass is
over 300~MeV.
Earlier studies of chiral extrapolations of lattice data for hadron masses
\cite{MASS}, magnetic moments \cite{MAGMOM} and charge radii \cite{RADII}
have shown that for quark masses above 50--60~MeV, hadron properties
behave very much as one would expect in a constituent quark model, with
relatively slow, smooth behavior as a function of the quark mass.
However, for $m_q \alt 50$~MeV one typically finds the rapid, nonlinear
variation expected from the nonanalytic behavior of Goldstone boson loops
\cite{CHIPT}.
The transition occurs when the pion Compton wavelength becomes larger than
the pion source --- essentially, the size of the extended nucleon.

Following the earlier work on chiral extrapolations of physical
observables, we expand the moments $\langle x^n \rangle_q$ at small
$m_\pi$ as a series in $m_\pi^2$.
Generally the expansion coefficients are (model-dependent) free
parameters.
On the other hand, the pion cloud of the nucleon gives rise to unique
terms whose nonanalyticity in the quark mass arises from the infra-red
behavior of the chiral loops.
Hence they are generally model independent.
In fact, the leading nonanalytic (LNA) term for the $u$ and $d$
distributions arising from a one-pion loop behaves as \cite{TMS}:
\begin{equation}
\langle x^n \rangle_q^{\rm LNA}
\sim m_\pi^2 \log m_\pi\ .
\label{eq:lna}
\end{equation}
Experience with the chiral behavior of masses and magnetic moments shows
that the LNA terms alone are not sufficient to describe lattice data for
$m_\pi \agt 200$~MeV \cite{MASS,MAGMOM}, so that extrapolation of lattice
data to $m_\pi \sim 0$, through the chiral transition region, requires a
formula which is consistent with both the heavy quark and chiral limits
of QCD.

In order to fit the lattice data at larger $m_\pi$, while preserving
the correct chiral behavior of moments as $m_\pi \to 0$, the moments
of $u-d$ are fitted with the form:
\begin{equation}
\label{eq:fit}
\langle x^n \rangle_{u-d}
=\ a_n\ +\ b_n\ m_\pi^2\
+\ a_n\ c_{\rm LNA}\ m_\pi^2
   \ln \left( \frac{m_\pi^2}{m_\pi^2 + \mu^2} \right)\ ,
\end{equation}
where the parameters $a_n$ and $b_n$ are {\em a priori} undetermined, and
the mass $\mu$ essentially determines the scale above which Goldstone
boson loops no longer yield rapid variation --- typically at scales
$\sim 500$~MeV.
(In fact, the mass $\mu$ corresponds to the upper limit of the momentum
integration if one applies a sharp cut-off in the pion loop integral
\cite{FF}.)
The coefficient $c_{\rm LNA} = -(3 g_A^2 + 1)/(4\pi f_\pi)^2$ has been
calculated in chiral perturbation theory \cite{CHPT}.
In the limit $m_\pi \to 0$ the form in Eq.(\ref{eq:fit}) is therefore
the most general expression for moments of the PDFs at
${\cal O}(m_\pi^2)$ which is consistent with chiral symmetry.
At larger $m_\pi$ values, where chiral loops are suppressed, the argument
of the logarithm in Eq.(\ref{eq:fit}) ensures that the effects of this
term are switched off.

Having motivated the functional form of the extrapolation formula, we
next apply Eq.(\ref{eq:fit}) to the lattice data from
Refs.\cite{QCDSF96,QCDSF97,BEST,MIT}.
In principle, Eq.(3) is only strictly applicable to full (unquenched) QCD,
and quenched chiral perturbation theory should be used to extrapolate
quenched data at small quark masses where the effects of pion and spurious
$\eta$ loops will dominate the $m_{\pi}$ dependence.
However, at the large quark masses where lattice calculations are
currently performed, chiral effects are strongly suppressed and,
as shown in Fig. 1, quenched and unquenched results are statistically
indistinguishable and have therefore been combined to improve overall
statistics.

While the current lattice data are at values of $m_\pi$ too high to
display any deviation from constituent quark behavior, it is not
{\em a priori} obvious why a lowest order form should be able to fit
data at $m_\pi \sim 1$~GeV.
Hence, it is useful to note that studies based on chiral quark models
suggest that Eq.(\ref{eq:fit}) can indeed provide a very good
parameterization of the $m_\pi$ dependence of PDF moments.
We illustrate this by taking a simple meson cloud model of the nucleon,
based on the MIT bag with pion cloud corrections introduced perturbatively
in an expansion in the infinite momentum frame \cite{IMF} about `bare'
nucleon states --- analogously to the cloudy bag model (CBM) \cite{CBM}.
Earlier studies of the $N$ and $\Delta$ masses \cite{MASS} and the
nucleon magnetic moments \cite{MAGMOM} established that the CBM gives
a good description of the lattice data over a wide range of quark mass.
The details of structure function calculations in the meson cloud model
are well known and can be found in the literature \cite{CLOUD,BOROS}
(see also \cite{LONG_VERSION}).
Since the model is not our main focus here, we simply show the results
for the $n=1, 2$ and 3 moments (for a bag radius of 0.8~fm and a $\pi NN$
dipole vertex form factor mass of 1.3~GeV \cite{CLOUD,BOROS}).
These are denoted in Fig.~1 by the small squares, and the $\chi^2$ fits to
these using the form (\ref{eq:fit}) are represented by the dashed curves
through them.
Clearly, Eq.(\ref{eq:fit}) provides an excellent fit to model data,
which are also in qualitative agreement with the calculated lattice
moments.
These results give us confidence that a fit to the lattice data based
on Eq.(\ref{eq:fit}) should be reasonable.

The results of the best $\chi^2$ fit to the lattice data for each moment
are given by the central solid lines in Fig.~1.
The inner envelopes around these curves represent fits to the extrema of
the error bars.
For the central curves, the value of the mass parameter $\mu$ that is most
consistent with all experimental moments is $\mu=550$~MeV.
This value of $\mu$ is comparable to the scale at which the behavior
found in other observables, such as magnetic moments and masses, switches
from smooth and constituent quark-like (slowly varying with respect to
the current quark mass) to rapidly varying and dominated by Goldstone
boson loops.
The similarity of these scales for the various observables simply
reflects the common scale at which the Compton wavelength of the pion
becomes comparable to the size of the hadron (without its pion cloud).
We also note that this is similar to the scale predicted by the
$\chi^2$ fits to the meson cloud model in Fig.~1.

At present, all of the lattice data are in a region where the moments
show little variation with $m_\pi^2$.
This, together with the relatively large errors, means that one cannot
distinguish between a linear extrapolation and one that includes the
correct chiral behavior, as Fig.~1 illustrates.
Consequently, it is not possible to determine $\mu$ from the current
lattice data.
In fact, with the current errors it is possible to consistently fit
both the lattice data and the experimental values with $\mu$ ranging
from $\sim 400$~MeV to 700~MeV.
The dependence on $\mu$ is illustrated in Fig.~1 by the difference
between the inner and outer envelopes on the fits.
The former are the best fits to the lower (upper) limits of the error
bars, while the latter use $\mu=450$ (650)~MeV instead of the central
value of $\mu=550$~MeV.
Data at smaller quark masses are therefore crucial to constrain this
parameter and guide an accurate extrapolation.

These results have significant implications for lattice calculations.
Unlike heavy quark systems, where it may be acceptable to work in a
reasonably small volume, calculations of the nucleon require an
accurate representation of the pion cloud.
Hence the volume must be sufficiently large that the pion Compton
wavelength of a reasonably light pion fits well within the volume.
Even though one need not calculate at the physical pion mass, the pion
must be light enough that the parameters of a systematic chiral
extrapolation are well determined statistically.
Specifically, from Fig.~1 it is clear that 5\% measurements down to
$m_{\pi}^2 = 0.05$~GeV$^2$ (requiring a spatial volume of order
(4.3~fm)$^3$) would provide data for an accurate chiral extrapolation.
This will require Terascale calculations \cite{LHPC}, first in the
quenched approximation with chiral fermions and eventually in full QCD,
which is necessary to produce the full pion cloud and the correct chiral
behavior embodied in the leading nonanalytic structure.
While this is demanding, with the hybrid Monte Carlo algorithm of
Ref.\cite{LIPPERT} requiring 8 Teraflops-years for full QCD, since the
total computational cost of this algorithm varies as $m_{\pi}^{-7.25}$,
we note that this is still a factor of 26 less than necessary for a brute
force evaluation at the physical quark mass.
Indeed, the discovery reported here brings reliable calculations of
hadronic properties within the capability of the next generation of
computers which will be available in the next 2--3 years.

In summary, we have investigated the quark mass dependence of moments
of quark distribution functions, with emphasis on both the physics in
the chiral limit and the scale at which the pion Compton wavelength
corresponds to the intrinsic size of the nucleon.
We proposed a low order formula for the $m_\pi$ dependence of moments,
which embodies the leading nonanalytic behavior expected from the chiral
properties of QCD, and used it to extrapolate the available lattice data
to the physical region.
The applicability of a low order expansion for the lattice data is also
supported by phenomenological chiral quark model studies.
Compared with linear extrapolations, which drastically overestimate the
experimental values, we find that within the current errors there is no
evidence of a discrepancy between the lattice data and experiment once the
correct dependence on quark mass near the chiral limit is incorporated.
This observation resolves an important long-standing problem with
{\em ab initio} calculations of hadron structure in QCD which has
persisted for more than a decade.
It not only removes a serious threat to the credibility of current
lattice calculations, but also provides the foundation for quantitative
calculation of hadron observables with the next generation of Terascale
computers.

\acknowledgements

We would like to thank C.~Boros, D.~Dolgov, R.~Edwards, R.~Horsley,
C.~Michael and D.~Richards for helpful discussions and communications,
and X.~Ji and M.~Savage for helpful discussions about the results in
Ref.\cite{CHPT}.
This work was supported by the Australian Research Council, the U.S.
Department of Energy contract \mbox{DE-AC05-84ER40150}, under which the
Southeastern Universities Research Association (SURA) operates the
Thomas Jefferson National Accelerator Facility (Jefferson Lab), and
cooperative research agreement \mbox{DE-FC02-94ER40818}.

\begin{center}
\begin{figure}
\epsfig{file=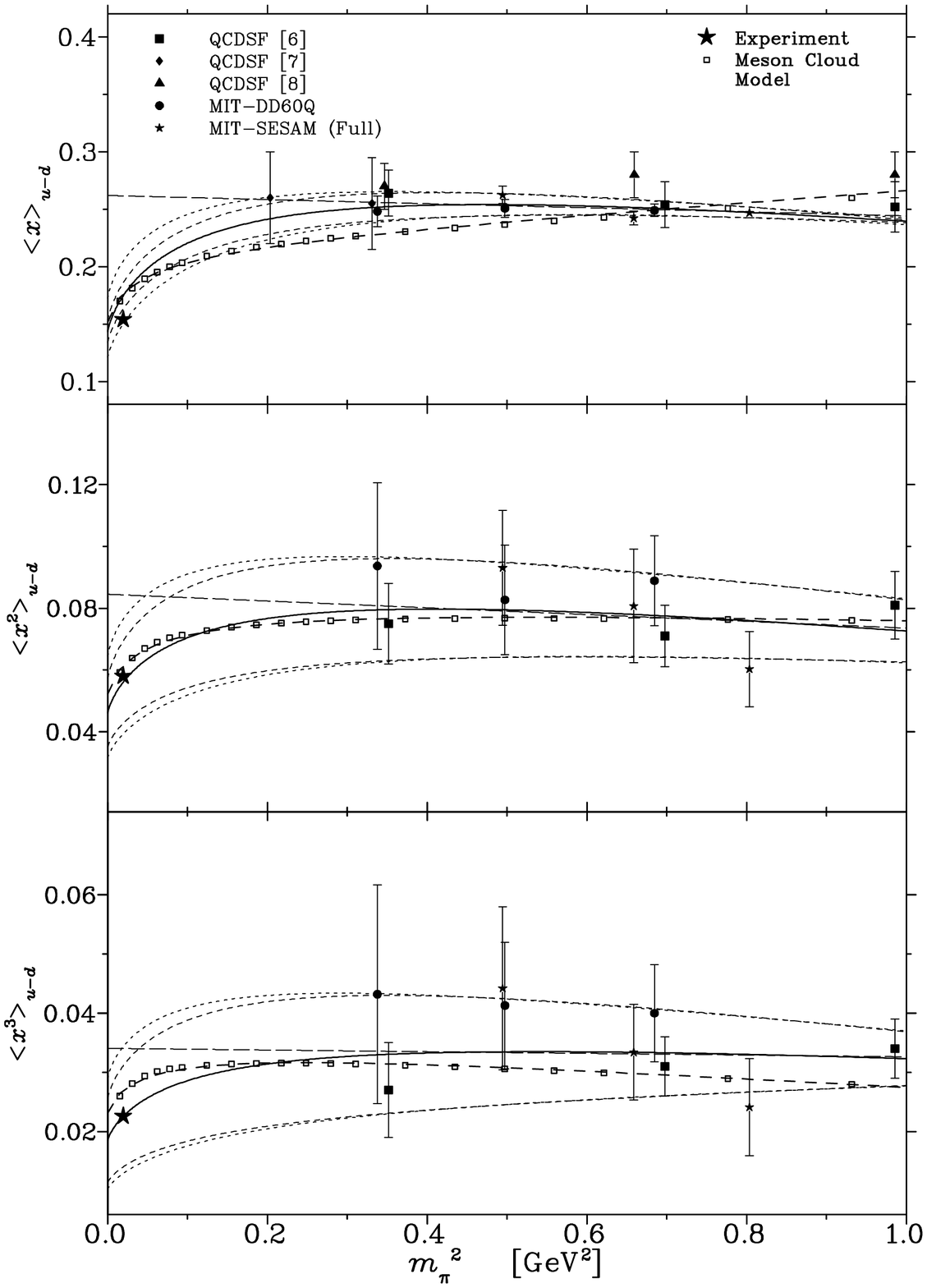,height=16cm} \vspace*{1cm}
\caption{Moments of the $u - d$ quark distribution.
  The straight (long-dashed) lines are linear fits to the data, while
  the curves have the correct LNA behavior in the chiral limit.
  For each moment, the best fit to the lattice data using
  Eq.(\protect\ref{eq:fit}) is shown by the solid curve (with
  $\mu=550$~MeV), while the inner envelope about this represents the
  statistical errors in the data.
  The best fit parameters are:
  $a_1=0.1427$,\, $b_1=-0.0624\,{\rm GeV}^{-2}$,\,
  $a_2=0.0459$,\, $b_2=-0.0245\,{\rm GeV}^{-2}$,\,
  $a_3=0.0184$,\, $b_3=-0.00666\,{\rm GeV}^{-2}$,
  which give a $\chi^2$ per degree of freedom of
  0.98, 0.60 and 0.60 for $n=1$, 2 and 3, respectively.
  The effect of the uncertainty in the parameter $\mu$ is illustrated
  by the outer lower (upper) short-dashed curves, which correspond to
  $\mu=450$ (650)~MeV.
  The small squares are the meson cloud model results
  \protect\cite{BOROS}, and the dashed curve through them best fits using
  Eq.(\protect\ref{eq:fit}).
  The star represents the phenomenological values taken from NLO fits
  \protect\cite{PARAMS} in the $\overline{\rm MS}$ scheme.}
\end{figure}
\end{center}

\end{document}